\documentclass[12pt]{article}

\pdfoutput=1
\usepackage{color}
\usepackage{epsfig,palatino}
\usepackage{mathrsfs}
\usepackage{ae} 
\usepackage[T1]{fontenc}
\usepackage[ansinew]{inputenc}
\usepackage{amsmath}
\usepackage{amssymb}
\usepackage{graphicx}
\usepackage{ulem}
\usepackage{xcolor}
\definecolor{darkblue}{cmyk}{0.9,0.9,0,0}
\usepackage[colorlinks=true,linkcolor=darkblue,citecolor=darkblue,urlcolor=darkblue]{hyperref}
\usepackage{cite}
\usepackage{wasysym}
\usepackage{varioref}
\usepackage[english]{babel}
\usepackage{tikz}

\usepackage[font={small}]{caption}

\usepackage{microtype}
\usepackage{todonotes}

\makeindex

\newcommand{\dd}[1][]{\mathrm{d}^{#1}}

\newcommand{\defas}{\mathrel{\mathop:}=}
\newcommand{\U}{\mathcal{U}}
\newcommand{\F}{\mathcal{F}}
\newcommand{\sdc}{\omega}
\newcommand{\ZZ}{\mathbb{Z}}
\newcommand{\NN}{\mathbb{N}}
\newcommand{\SP}{\alpha}
\newcommand{\mzv}[2][]{\zeta^{#1}_{#2 }}
\newcommand{\motivic}{\mathfrak{m}}
\newcommand{\deRham}{\mathfrak{dr}}
\newcommand{\Fourier}[1]{\mathcal{F}\left[ #1 \right]}

\newcommand{\Azurite}{\href{https://bitbucket.org/yzhphy/azurite/}{\textsc{Azurite}}}
\newcommand{\FIRE}{\href{https://bitbucket.org/feynmanIntegrals/fire/}{\textsc{FIRE}}}
\newcommand{\HyperInt}{\href{http://bitbucket.org/PanzerErik/hyperint/}{\texttt{\textup{HyperInt}}}}
\newcommand{\Mathematica}{\textit{Mathematica}}

\title{Five-loop massless propagator integrals}

\begin{document}

\thispagestyle{empty}

\renewcommand{\thefootnote}{\fnsymbol{footnote}}
\setcounter{page}{1}
\setcounter{footnote}{0}
\setcounter{figure}{0}


\noindent

\hfill
\begin{minipage}[t]{35mm}
\begin{flushright}
UUITP-02/18 \\
TCDMATH 18-02
\end{flushright}
\end{minipage}

\vspace{1.0cm}

\begin{center}
{\Large\textbf{\mathversion{bold}
Five-loop massless propagator integrals\\
}\par}

\vspace{1.0cm}

\textrm{Alessandro Georgoudis\textsuperscript{1,2}, Vasco Goncalves\textsuperscript{2}, Erik Panzer\textsuperscript{3}, Raul Pereira\textsuperscript{1,4}}
\\ \vspace{1.2cm}
\footnotesize{\textit{
\textsuperscript{1}Department of Physics and Astronomy, Uppsala University
Box 516, SE-751 20 Uppsala, Sweden\\
\textsuperscript{2}ICTP South American Institute for Fundamental Research, IFT-UNESP, S\~ao Paulo, SP Brazil 01440-070\\
\textsuperscript{3}All Souls College, University of Oxford, OX1 4AL, Oxford, UK \\
\textsuperscript{4}School of Mathematics and Hamilton Mathematics Institute, Trinity College Dublin, Dublin, Ireland
}  
\vspace{4mm}
}

\end{center}

\begin{abstract}
We develop a method to obtain $\epsilon$-expansions of massless two-point integrals in position space, based on the constraints implied by symmetries of the asymptotic expansion of conformal four-point integrals.
Together with parametric integration, we are able to fix the expansions of 170 genuine five-loop master integrals. In particular, we computed the expansions of all planar master integrals up to transcendental weight 9.
\end{abstract}
\noindent

\setcounter{page}{1}
\renewcommand{\thefootnote}{\arabic{footnote}}
\setcounter{footnote}{0}

\setcounter{tocdepth}{2}

\def\nref#1{{(\ref{#1})}}

\newpage

\parskip 5pt plus 1pt   \jot = 1.5ex

\newpage

\parskip 5pt plus 1pt   \jot = 1.5ex



\section{Introduction}
\label{sec:intro}
Increasingly high precision measurements at particle colliders like LEP and LHC call for the calculation of higher order perturbative corrections to various physical observables. In order to have a complete interplay between experiment and perturbative quantum field theory, it is thus crucial to evaluate Feynman integrals with several scales and loop momenta. The complexity of this task grows significantly as the number of loop momenta or scales increases. 

Through Integration By Parts (IBP) identities \cite{Chetyrkin:1981qh} one can, in principle, reduce the determination of infinitely many integrals of certain families to the evaluation of a finite set of master integrals \cite{SmirnovPetukhov:Finite}. These identities have been efficiently implemented in several publicly available codes \cite{Anastasiou:2004vj,Lee:2013mka,Smirnov:2014hma,vonManteuffel:2012np,Ueda:2016yjm,Maierhoefer:2017hyi}, but such reductions still constitute a bottleneck in several applications; particularly at high loop orders.

In this work we will focus on the determination of massless propagator-type Feynman integrals with just one external momentum, which are usually denoted as $p$-integrals.  These integrals have many applications:  $l$-loop p-integrals can be used to compute the $(l+1)$-loop counterterms of any Feynman integral \cite{Chetyrkin:1984xa}, to obtain coefficient functions in deep inelastic scattering, to give boundary conditions for form factor computations using the differential equation approach\footnote{See \cite{Baikov:2015tea,Henn:2013nsa} for a review on this subject.} or to compute series expansions of multi-scale integrals through the method of asymptotic expansions.

All four-loop master $p$-integrals have been computed in \cite{Baikov:2010hf,Lee:2011jt} to transcendental weights $7$ and $12$ respectively, while three-loop $p$-integrals were determined in \cite{Chetyrkin:1980pr}. 
In this paper, we present $\epsilon$-expansions for 170 of the 187 genuine five-loop master integrals for two-point functions in position space, including all planar integrals up to transcendental weight 9. By planar duality, the latter also provide the $\epsilon$-expansions of all planar $p$-integrals (in momentum space).
Furthermore, by performing the Fourier transform, we were also able to find the leading order in the $\epsilon$-expansion of twenty non-planar master $p$-integrals. 
Most of our results were obtained by analyzing the asymptotic expansions of finite conformal integrals with three external off-shell legs, while some further masters were evaluated with the program {\HyperInt} \cite{Panzer:2014caa}. 

The results obtained are presented in ancillary files, which we explain in the {\Mathematica} notebook \texttt{FiveLoopMasters.nb}.

In section~2 we explain how to use the program {\Azurite} \cite{Georgoudis:2016wff} to find the set of master five-loop two-point integrals in position space. In general these integrals are divergent in $d=4$, so we use dimensional regularization where $d=4-2\epsilon$ and divergences show up as poles in $\epsilon$. Then in section~\ref{sec:bootstrap} we describe how to exploit the conformal symmetry of multi-scale integrals to constrain the expansions of the master integrals. Several checks of the results are discussed in section~\ref{sec:checks} and we conclude with some general remarks on the structure of the $\epsilon$-expansions. In the appendix we illustrate how the Fourier transform relates non-planar integrals from position to momentum space.

\section{Finding the master integrals}
In order to determine master integrals we need to choose a basis of propagators. In the case in study, the basis is represented by propagators of the form:
\begin{equation}\label{eq:basex}
	x_{ij}=x_i-x_j\,.
\end{equation} 
We consider diagrams with five internal vertices, labeled by indices $\{1,2,3,4,5\}$, and two external points, labeled by indices $\{0,6\} $. By translation invariance of \eqref{eq:basex}, we can always move one external point to the origin $x_0=0$ and may then rescale all points so that the other external vertex $y \defas x_6$ has norm $y^2=1$. 
Our family of integrals can then be written as
\begin{equation}\label{eq:pint}
	I(a)
	=
	\int \prod_{k=1}^5 \dd[d] x_k 
	\prod\limits_{0\le i < j \le 6}\frac{1}{x_{ij}^{2a_{ij}}},
\end{equation}
where the $20$ indices $a_{ij}$ are understood to take integer values, and the integration is over the position of the internal vertices in $d=4-2\epsilon$ dimensional space (in the sense of dimensional regularization).
We can represent every such integral through its associated Feynman graph, where propagators $x_{ij}^2$ are encoded as edges connecting the vertices $i$ and $j$. For example,
\begin{equation*}
	\int \frac{\dd[d] x_1 \dd[d]x_2 \dd[d] x_3 \dd[d] x_4 \dd[d] x_5}{x_{01}^2 ~ x_{12}^2 ~ x_{23}^2 ~ x_{25}^2 ~x_{36}^2 ~x_{34}^2 ~x_{14}^2 ~x_{45}^2 ~x_{56}^2 }=
  \begin{tikzpicture}[baseline={([yshift=-1.8ex]current bounding box.center)},scale=0.3,node/.style={draw,shape=circle,fill=black,scale=0.4}]
    \path (-4,0) node[label=\small $0$] (p0) {}
(0,0) node[label=\small $x_1$] (p1) {}
(4,4) node[label=\small $x_2$] (p2) {}
(10,4) node[label=\small $x_3$] (p3) {}
(14,0) node[label=\small $y$] (p4) {}
(10,-4) node[label=\small $x_5$] (p5) { }
(4,-4) node[label=\small $x_4$] (p6) { };
\draw[ thick]  (-4,0) -- (0,0) -- (4,4) -- (10,4) --( 14,0) -- (10,-4) -- (4,-4) -- (0,0);
\draw[ thick] (4,-4) -- (10,4);
\draw[ thick] (4,4) -- (10,-4);
  \end{tikzpicture} \,.
\end{equation*}
The integrals $I(a)$ can be classified according to which denominators appear in the integrand \eqref{eq:pint}. Each of the $2^{20}$ subsets of denominators thus determines a family of integrals, called a topology.
Luckily, not all topolgies need to be analysed:
	\begin{itemize}
		\item We can discard topologies in which a tadpole appears, since they evaluate to zero in dimensional regularization.

		\item Topologies which are factorizable can be skipped as well, as they are expressed as a product of known sub integrals with fewer loops.

		\item The global symmetry $S_2 \times S_5$ of \eqref{eq:pint} identifies topologies that only differ in the labelling of the vertices.
	\end{itemize}%
We are then left with 3227 topologies that need to be analysed.
For this purpose, we used the program {\Azurite} \cite{Georgoudis:2016wff} to obtain the list of five-loop master integrals for our choice of denominators.
The program works by considering the integration by parts identities (IBPs) on a single topology at a time.
These identities follow from the fact that the integral of a total derivative is vanishing:
\begin{equation*}
	\int \left( \prod_{k=1}^5 \dd[d] x_k  \right) 
	\frac{\partial }{\partial x_r^{\mu}}
	\left(
		P^{\mu}
	\prod\limits_{0\le i < j \le 6}\frac{1}{x_{ij}^{2a_{ij}}}
	\right)
	=0\,,
\end{equation*}
where the (Lorentz) vector $P^{\mu}$ is a polynomial in internal and/or external points, and $1\leq r \leq 5$.
Azurite further constricts the relations to contain only integrals from a single topology. These relations between integrals with different sets of exponents can be used to reduce any integral $I$ to a linear combination of finitely many master integrals $M_k$:
\begin{equation*}
	I(a) = \sum_{k} c_k(a,\epsilon) \,M_k(\epsilon) \,.
\end{equation*}	
Since the relations are solved with numerical coefficients and on a single topology, this approach is very suitable for our situation.

If a given topology contains several master integrals, the elements of the basis depend on several choices made during the reduction process, such as the relative order chosen for the propagators in the integrals.
Finally, we find that there are 187 master integrals from 154 topologies of genuine five-loop integrals (not counting master integrals that factor into integrals with fewer loops).

\subsection{Planar Masters}
Restricting to the planar sector, we find 95 non-factorizable planar master integrals from 92 planar topologies. In order to complete the basis of planar master integrals, we have to add those 23 master integrals that can be constructed as a product of lower loop masters.

\begin{figure}[h]
\centering
  \begin{tikzpicture}[scale=0.5,node/.style={draw,shape=circle,fill=black,scale=0.4}]
\draw[ thick]  (-4,0) -- (0,4) -- (4,0) -- (0,-4) --(-4,0);
\draw[ thick] (-4,0) -- (0,2)--(2,0)--(0,-2)--(-4,0);
\draw[ thick] (0,4) -- (0,2);
\draw[ thick] (4,0) -- (2,0);
\draw[ thick] (0,-4) -- (0,-2);
  \path (-4,0) node[label=\small $0$] (p0) {}
(4,0) node[label=\small{$y$}] (p1) {}
(8.5,0) node[label=\small $p$] (p2) {}
(20.5,0) node[label=\small $p$] (p3) {};
\draw[thick,->] (5,0)--(7,0);
\draw[ thick] (8,0)--(21,0)--(19,0)--(17.5,3*0.866)--(16,0)--(14.5,3*0.866)--(13,0)--(11.5,3*0.866)--(10,0);
\draw[ thick] (11.5,3*0.866)--(17.5,3*0.866);
\end{tikzpicture}
	\caption{Duality transformation for a planar five-loop two-point integral.}
	\label{fig:Duality}%
\end{figure}
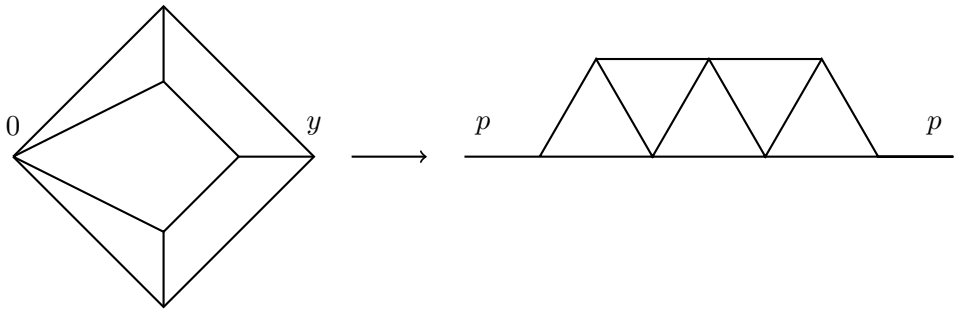

Recall that, in the planar sector, we can identify integrals in position space with integrals in momentum space.
Explicitly, considering $x_i$ as loop momenta $k_i$ and $x_6$ as the external momentum $p$, the denominators $x_{ij}$ can be seen as the momentum space denominators
\begin{equation*}
	k_i^2, \qquad  
	\left( k_i - p \right)^2  \quad\text{and}\quad
	\left( k_i - k_j \right)^2 \quad \text{with $i<j$.}
\end{equation*}
Diagrammatically, this identification between planar position- and momentum-space integrals just amounts to the well-known planar duality, which is depicted in Figure \ref{fig:Duality}.

\section{Bootstrap of master integrals}
\label{sec:bootstrap}

There are several methods for the evaluation of master $p$-integrals. At four loops, all master integrals were obtained up to transcendental weight 12 by considering dimensional recurrence relations and analyticity considerations \cite{Lee:2011jt}.
In principle, this approach should also work at five loops, though it is unclear to us if the tremendous increase in complexity can be dealt with in practice.
The cut-and-glue method introduced in \cite{Baikov:2010hf} could also be used at five loops. However, at four loops it was necessary to consider one hundred five-loop vacuum diagrams, so we expect this number would increase dramatically at five loops.

On the other hand, if the master integrals are linearly reducible, one can try to evaluate them directly by integrating out Schwinger parameters with {\HyperInt}. This has proved a powerful method in many applications, but in practice, at the very high complexity of the integrals in question, the calculations can become extremely time- and memory consuming. In particular, the evaluation of divergent integrals to higher orders in $\epsilon$ (sometimes needed to reach transcendental weight $9$) can be prohibitively challenging.

Another approach is to integrate directly in position space, which is extremely efficient and successful in the language of graphical functions \cite{Schnetz:GraphicalFunctions}. Exploiting the theory of single-valued generalizations of hyperlogarithms, these tools can be used to compute $\epsilon$-expansions of very complicated integrals \cite{Schnetz:NumbersAndFunctions}.
 
In what follows we present a method that allows us to obtain the $\epsilon$-expansions of most master integrals without any explicit integration. The idea is to impose several constraints which translate into a linear system of equations on the coefficients of the $\epsilon$-expansions. 
At the end of this section, we explain how parametric integration can be used to obtain the rest of our results.

\subsection{Constraints from IBP identities}
 	
It turns out that the reduction to master integrals also contains crucial information about the expansions of the masters themselves.\footnote{%
		This observation is not entirely new and it has been explored before in \cite{Czakon:2004bu,Chetyrkin:2006dh}.
}
First note that the degree of divergence of an $l$-loop two-point integral should not exceed $l$. This was shown for momentum-space integrals in \cite{Chetyrkin:1984xa}, and is expected to hold in position space upon consideration of a coordinate space version of the $R^*$-operation \cite{Eden:2012fe}.
On the other hand, the coefficients $c_k(\epsilon)$ that appear in the IBP reduction to master integrals might also have poles in $\epsilon$. If the highest pole of the coefficients $c_k(\epsilon)$ in $\epsilon$ is of order $m>0$, then the integral has an apparent divergence of order $5+m$,
\begin{equation*}
 	I(a)=\sum_{j=0}^\infty \alpha_j \,\epsilon^{-5-m+j}\,,
\end{equation*}
where the coefficients $\alpha_j$ are determined by the $c_k$ as certain linear combinations of the coefficients of the $\epsilon$-expansions of the master integrals.
All the poles of order higher than five are necessarily spurious, and we obtain a set of linear constraints 
\begin{equation}\label{IBpconstraints}
	\alpha_j=0, \quad \text{for}\quad 0 \leq j < m.
\end{equation}
 In general this is a homogeneous system of equations, so one would hope at best to find some vanishing coefficients in the expansions of master integrals. However, several of the masters are products of lower-loop integrals, which are all known, so the system of equations becomes non-homogeneous and it is possible to obtain several non-trivial coefficients in the expansions. 
 	
All IBP reductions were performed with the  C++ version of the program {\FIRE}	\cite{Smirnov:2014hma}. The two-point integrals we consider as seeds for this system of equations are the ones obtained through the expansion of conformal integrals described below. Furthermore, we keep all relations obtained, even for seeds whose reductions do not involve any product topologies, as these still reduce the number of undetermined coefficients.

\subsection{Constraints from conformal integrals}
	 
While the considerations above are straightforward to implement, they provide far too few constraints to fix all required terms in the $\epsilon$-expansions of the master integrals.
To obtain further relations, we exploit that two-point integrals appear in the asymptotic expansions of multi-scale integrals. While considering more complicated integrals might seem counterproductive at first, it is crucial to observe that symmetries of those integrals are still present in their asymptotic expansions. In this situation, we can deduce constraints on these expansions without knowing the value of the multi-scale integral itself.
	
For example, let us consider conformal four-point integrals. These are position space integrals with propagators $x_{ij}$, just as in \eqref{eq:pint}, but with four (instead of two) external vertices $x_1,\dots,x_4$ (we label the internal vertices $x_5,\dots,x_9$). Conformal invariance means that, up to a simple rational prefactor, these integrals only depend on two cross-ratios, which we may choose as
\begin{equation}
	u = \frac{x_{12}^2 x_{34}^2}{x_{13}^2 x_{24}^2}
	\quad\text{and}\quad
	v = \frac{x_{14}^2 x_{23}^2}{x_{13}^2 x_{24}^2}
	.
	\label{eq:uv-4pt}%
\end{equation}
While such conformal integrals are difficult to evaluate even at four loops\footnote{%
	The position-space analogue of the number of loops in momentum space is the number of internal vertices.
}
\cite{Eden:2016dir}, it is possible to extract a coincidence limit through the method of asymptotic expansions. Using a conformal transformation we are free to set $x_1$ to the origin and send $x_4$  to infinity, in which case the coincidence limit $x_2 \to x_1=0$ leaves us with two distinct scales, $x_{2}^2 \ll x_{3}^2$. 
We can then approximate the five-loop conformal integral by separating each integration into two regions: one where the integration variable $x_i$ is of the order of $x_{2}$ and the other where it is of the order of $x_{3}$.
For example, if the integration variables are large, we can approximate the denominators by
\begin{equation}
	\frac{1}{(x_2 - x_i)^2} 
	= \sum_{n=0}^\infty \frac{(2 x_2 \cdot x_i - x_2^2)^n}{(x_i^2)^{n+1}} 
	\quad (\text{if $x_2^2< x_i^2$}).
	\label{eq:propagator-expansion}%
\end{equation}
We can see that in each region of the asymptotic expansions, the conformal integral becomes a product of two-point integrals, where the external points are either $x_1$ and $x_2$, or $x_1$ and $x_3$. The full coincidence limit of the five-loop conformal integral is given by the sum over 32 possible such regions.

The last step in the method of asymptotic expansions is to ignore factorization scales and integrate over the full domain. In order for this to be valid, it is crucial to consider dimensionally regularized integrals, and require that scaleless integral vanish \cite{Beneke}. Let us also note that a graphical formula for the coincidence limit has been proposed in \cite[Conjecture~4.12]{Schnetz:NumbersAndFunctions}.
	
If we reduce the two-point integrals obtained, we are able to express the coincidence limit of the four-point integral through products of $k$ and $(5-k)$-loop master integrals. By further substituting the expansion of lower-loop master integrals, we obtain the expansion of the conformal integral as a linear combination of the five-loop master integrals we wish to obtain. At this point it is crucial to remember that if the original integral is conformal, then its expansion must be expressed through the two cross-ratios
\begin{equation}
	u=\frac{x_{2}^2}{x_{3}^2} 
	\,, \qquad\qquad 
	v=\frac{x_{23}^2}{x_{3}^2}\,.
	\label{eq:uv-3pt}%
\end{equation}
Meanwhile, the asymptotic expansion introduces a dependence on the scales of the two-point integrals: $x_{2}^2$ and $x_{3}^2$. The cross-ratio dependence would be restored in four dimensions, but we are considering dimensionally regularized integrals, so that the master integrals are given as expansions in $\epsilon$, and after all terms have been combined, the coincidence limit of the four-point integral must be finite and independent of any of the spurious scales.
	
To be more precise, the asymptotic expansion of each five-loop four-point integral at lowest order in $u$ and $1-v$ has an expansion of the form (up to terms that vanish as $\epsilon \rightarrow 0$)
\begin{equation}
	\sum_{n=0}^5 \sum_{k=0}^n \sum_{l=0}^{n-k} 
	\beta_{nkl} \, \log^k(x_{3}^2) \log^l(u)  \epsilon^{-5+n} \,,
\end{equation}
where the coefficients $\beta_{nkl}$ are linear combinations of the $\epsilon$-expansion coefficients of the five-loop master integrals. The coincidence limit of a convergent conformal integral must be finite in $\epsilon$ and depend only on the cross-ratios, so we obtain the following set of linear constraints 
\begin{equation}\label{FiniteConstraints}\begin{split}
	\beta_{nkl}&=0 \,,  \qquad\text{where $n<5$ and $k\geq 0$,}\\
	\beta_{5kl}&=0 \,,  \qquad\text{for $k> 0$.}
\end{split}\end{equation}

For each four-point integral we can still permute the external points, so that taking $x_{12}$ to zero corresponds to different limits of the conformal integrals. The permutations of external points lead to six transformations of the cross ratios $(u,v)$:
\begin{equation}\begin{split}
	(u,v)     &\approx (u/v, 1/v) \approx (0,1) \,,\\
	(v,u)     &\approx (1/v, u/v) \approx (1,0) \,,\\
	(1/u,v/u) &\approx (v/u, 1/u) \approx (\infty,\infty) \,.
\end{split}\end{equation}
We can see that for small $u$ and $1-v$ they organize in three pairs which correspond to three distinct regions in cross-ratio space. The elements of each pair correspond to different gauge choices of the same coincidence limit (we fix either $x_1$ or $x_2$ to be at the origin, while taking $x_{12}$ small in both cases), which means that the asymptotic expansions lead to different sets of coefficients $\beta_{nkl}$, which implies that a given limit can be expressed though different combinations of two-point integrals. This is quite important as it imposes constraints on the finite part of the two-point integrals, unlike the constraints in \eqref{IBpconstraints} or \eqref{FiniteConstraints}.
	
Conformal symmetry still imposes further restrictions that we can translate into contraints on the master integrals. The cross ratios \eqref{eq:uv-4pt} are invariant under the Klein four-group
\begin{equation}
	\{\text{id},(1\ 2)(3\ 4),(1\ 3)(2\ 4),(1\ 4)(2\ 3)\}
	\cong \ZZ_2 \times \ZZ_2
\end{equation}
of double transpositions of the four external points. While such relabellings leave an integral $I(u,v)$ unchanged as a function of the cross-ratios, these transformations can lead to different expressions (in terms of two-point integrals) for the asymptotic expansions in coincidence limits.
One can take this strategy further by applying it to any conformal four-point subintegral of the original five-loop integral, resulting in the so called magic identities for conformal integrals \cite{Drummond:2006rz}, which coincide with Schnetz's twist relations \cite{Schnetz:2008mp} for graphical functions \cite{Schnetz:GraphicalFunctions}.
	
\begin{figure}[t]
	\centering
	\begin{tikzpicture}[scale=0.45,node/.style={draw,shape=circle,fill=black,scale=0.5}]
	\path (-16.8,-0.5) node[label=\small $3$] () {}
	(-15.2,-.5) node[label=\small $7$] () {}
	(-12.5,-.5) node[label=\small $9$] () {}
	(-10.8,-.5) node[label=\small $8$] () {}
	(-9.2,-.5) node[label=\small $2$] () {}
	(-13,1.5) node[label=\small $1$] () {}
	(-13,-3.5) node[label=\small $5$] () {}
	(-7.8,-.5) node[label=\small $1$] () {}
	(-6.2,-.5) node[label=\small $6$] () {}
	(-2.8,-.5) node[label=\small $5$] () {}
	(-1.2,-.5) node[label=\small $2$] () {}
	(-4.5,1.5) node[label=\small $3$] () {}
	(-4.5,-3.5) node[label=\small $4$] () {}
	(7.8,-.5) node[label=\small $2$] () {}
	(6.2,-.5) node[label=\small $5$] () {}
	(2.8,-.5) node[label=\small $6$] () {}
	(1.2,-.35) node[label=\small $1$] () {}
	(4.5,1.5) node[label=\small $3$] () {}
	(4.5,-3.5) node[label=\small $4$] () {}
	(16.8,-0.5) node[label=\small $1$] () {}
	(15.2,-.5) node[label=\small $8$] () {}
	(12.5,-.5) node[label=\small $9$] () {}
	(10.8,-.5) node[label=\small $7$] () {}
	(9.3,-.5) node[label=\small $5$] () {}
	(13,1.5) node[label=\small $2$] () {}
	(13,-3.5) node[label=\small $3$] () {}
	(-3,0) node[fill=green,circle,scale=0.5]() {}
	(-13,-2) node[fill=green,circle,scale=0.5]() {}
	(9,0) node[fill=green,circle,scale=0.5]() {}
	(6,0) node[fill=green,circle,scale=0.5]() {};
	\draw[ thick]  (-17,0)--(-11,0)--(-13,2)--(-15,0)--(-13,-2)--(-13,2);
	\draw[ thick]  (-13,-2)--(-11,0)--(-9,0);
	\draw[ thick]  (-8,0)--(-3,0)--(-4.5,2)--(-6,0)--(-4.5,-2)--(-3,0)--(-1,0);
	\draw[ dashed, color=red]  (-8,0)to[out=-45,in=225](-3,0);
	\node at (0, 0) {$=$};
	\draw[ thick]  (17,0)--(11,0)--(13,2)--(15,0)--(13,-2)--(13,2);
	\draw[ thick]  (13,-2)--(11,0)--(9,0);
	\draw[ thick]  (8,0)--(3,0)--(4.5,2)--(6,0)--(4.5,-2)--(3,0)--(1,0);
	\draw[ dashed]  (1,0) to[out=-45,in=225](6,0);
	\draw[dashed, color=blue] (8,0) -- (4.5,2); 
	\end{tikzpicture}
	\caption{Example of a magic identity for a five-loop four-point integral. Dashed lines represent numerators, while blue and red colors represent square and cubic powers respectively. The power of the identity relies on the fact that one of the external points (marked with green) in the three-loop subintegral is an integration variable of the complementary two-loop subintegral.}
	\label{fig:Magic}%
\end{figure}
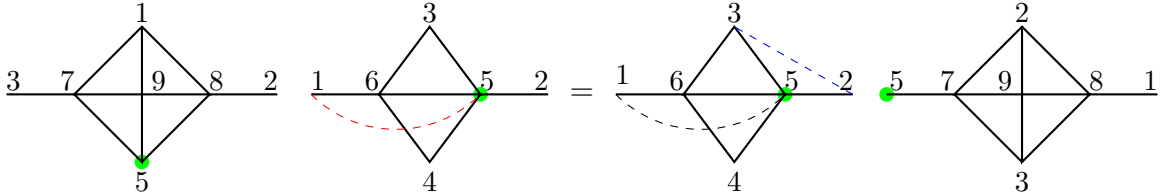

In this way we are able to generate classes of equal conformal integrals, whose asymptotic expansions will usually result in different linear combinations of two-point integrals, and might even depend on different sets of five-loop master integrals. 
For example, let us consider the following five-loop four-point integral
\begin{equation}
	I \defas \int\frac{ \dd[4] x_5  \dd[4] x_6 \dd[4] x_7 \dd[4] x_8 \dd[4] x_9 \cdot  x_{15}^6}{x_{16}^2 x_{17}^2 x_{18}^2 x_{19}^2  x_{25}^2 x_{28}^2 x_{35}^2 x_{36}^2 x_{37}^2 x_{45}^2 x_{46}^2  x_{56}^2 x_{57}^2 x_{58}^2 x_{59}^2   x_{79}^2 x_{89}^2} \,.
	\label{eq:twist-ex1}%
\end{equation}
One of its three-loop subintegrals is a four-point integral itself,
\begin{equation*}
	I_\text{sub}
	=\int \frac{\dd[4] x_7 \dd[4] x_8 \dd[4] x_9}{x_{17}^2 x_{18}^2 x_{19}^2 x_{28}^2   x_{37}^2 x_{57}^2 x_{58}^2 x_{59}^2 x_{79}^2 x_{89}^2} 
	= \frac{1}{x_{15}^6 x_{23}^2} \Phi(\tilde u , \tilde v)\,,
\end{equation*}
where the cross ratios $\tilde u$ and $\tilde v$ depend on the points $x_1$, $x_2$, $x_3$ and $x_5$. 
If we exchange $x_1$ with $x_2$ and $x_3$ with $x_5$ simultaneously, then the function of the cross ratios remains the same, while the factor carrying the conformal weights changes accordingly,
\begin{equation*}
I'_\text{sub}
= \int \frac{\dd[4] x_7 \dd[4] x_8 \dd[4] x_9}{x_{18}^2 x_{27}^2 x_{28}^2 x_{29}^2   x_{37}^2 x_{38}^2 x_{39}^2 x_{57}^2 x_{79}^2 x_{89}^2}
= \frac{1}{x_{15}^2 x_{23}^6} \Phi(\tilde u , \tilde v)\,.
\end{equation*}
By exchanging $I_\text{sub}$ with $I'_\text{sub}$ in 	\eqref{eq:twist-ex1} and multiplying by the factor $x_{23}^4/x_{15}^4$, we obtain a new conformal integral which is equal to $I$,
\begin{equation}
	I'\defas\int\frac{ \dd[4] x_5  \dd[4] x_6 \dd[4] x_7 \dd[4] x_8 \dd[4] x_9 \cdot  x_{15}^2 x_{23}^4}{x_{16}^2 x_{18}^2 x_{25}^2 x_{27}^2 x_{28}^2 x_{29}^2 x_{35}^2 x_{36}^2 x_{37}^2 x_{38}^2 x_{39}^2 x_{45}^2 x_{46}^2 x_{56}^2 x_{57}^2 x_{79}^2 x_{89}^2} \,.
	\label{eq:twist-ex2}%
\end{equation}
The equality $I=I'$ is non-trivial since the integrands have different numerators and a different number of propagators between internal points, as we can see in Figure \ref{fig:Magic}.
Hence, the asymptotic expansions of \eqref{eq:twist-ex1} and \eqref{eq:twist-ex2} in a coincidence limit will lead to different expressions in terms of two-point integrals.
	
For each element in an equivalence class of conformal four-point integrals, we can derive a set of equations like \eqref{FiniteConstraints}, which amounts to enforcing finiteness and absence of spurious scales. More importantly, the equality of conformal integrals implies further linear relations for the master two-point integrals. If the expansions of the integrals above in the coincidence limit $u\rightarrow 0$ are (at $\epsilon=0$)
\begin{equation*}
	I= \sum_{l=0}^{5} \beta_{50l} \,  \log^l(u)  \qquad\text{and}\qquad
	I'= \sum_{l=0}^{5} \beta'_{50l} \,  \log^l(u) \,,
\end{equation*}
then their coefficients must satisfy
\begin{equation}\label{MagicConstraints}
	\beta_{50l} = \beta'_{50l} \,,
	\qquad \text{for $0\leq l \leq 5$.}
\end{equation}

It is clear that the more conformal integrals we consider, the more powerful our bootstrap becomes, and a nice feature of the method is that magic identities and asymptotic expansions can be generated without much effort. The five-loop conformal integrals we have used in this bootstrap of master integrals have two different origins. On one hand we considered the generalized ladder diagrams introduced by Drummond \cite{Drummond:2012bg}
\begin{equation}
	L(\{a_i\})
	= \int \frac{1}{x_{15}^2}
	\prod_{i=5}^8\frac{\dd[4] x_i}{ x_{a_i,i}^2 x_{4,i}^2 x_{i,i+1}^2} 
	\frac{\dd[4] x_9}{x_{29}^2 x_{39}^2 x_{49}^2} \,,
\end{equation}
with the letters $a_i$ are taken from the set $\{2,3\}$. Apart from these, we consider the two hundred five-loop conformal integrals that appear in the four-point function of protected operators in $\mathcal{N}=4$ SYM \cite{Georgoudis:2017meq}.
The results obtained through the bootstrap and the parametric integration below were sufficient to fix the coincidence limits of all conformal integrals considered. In this way, explicit results for four-point integrals were completely unnecessary for the determination of the two-point master integrals. Instead, we used the few known four-point integrals as independent checks (see section~\ref{sec:checks}).

By solving the constraints obtained through IBP identities and conformal symmetry of four-point integrals we obtained relations for 169 of the 187 genuine five-loop master integrals, and they were sufficient to determine 98 of them to transcendental weight 9, 37 to weight 8, 19 to weight 7 and 4 to weight 6. Note that this counting includes both planar and non-planar master integrals in position space, and all but two of the 95 planar masters were fixed to transcendental weight 9 in this way. 

The attentive reader will wonder why the discussion was restricted to the first order in $u$ and $(1-v)$ of the asymptotic expansions. One could in principle recover more terms in the expansions of the conformal integrals, and by imposing finiteness, conformal symmetry and permutation symmetry one would derive a larger number of constraints on the master integrals. However, in order to obtain the subleading terms in the expansion of the conformal integrals one would have to reduce extremely complicated five-loop two-point integrals. The task of performing IBP reductions is a real bottleneck in this method, and even the terms appearing at leading order are at the boundary of feasibility.

\subsection{Parametric integration}

The method introduced above is extremely powerful in determining the $\epsilon$-expansions of massless two-point integrals, and it also provides relations between many of the undetermined coefficients.
In order to extract further data from those relations, we computed a few integrals directly. 
Recall that the integrals \eqref{eq:pint} also have a representation
\begin{equation}
	I(a) = \Gamma(\sdc) \left(\prod_i \int_0^{\infty}\frac{\SP_i^{a_i-1} \dd \SP_i}{\Gamma(a_i)} \right)
	\frac{\U^{\sdc-d/2}}{\F^{\sdc}}
	\label{eq:parametric}%
\end{equation}
in terms of graph polynomials $\U,\F \in \ZZ[\SP_1,\ldots,\SP_{20}]$ in the Schwinger parameters $\SP_i$, see \cite{Nakanishi:GraphTheoryFeynmanIntegrals,GolzPanzerSchnetz:GfParam}. Via integrating by parts,\footnote{%
	The boundary terms $\SP_i \rightarrow 0,\infty$ are supported only on a discrete set of values for $a_i$, hence while \eqref{eq:parametric-ibp} is literally true only on the dense, open complement of this set, both sides of the equation have the same analytic continuations in $a_i$.
}
\begin{equation}
	\int_0^{\infty} \frac{\SP_i^{a_i-1} \dd \SP_i}{\Gamma(a_i)} f(\SP)
	= \int_0^{\infty} \frac{\SP_i^{a_i} \dd \SP_i}{\Gamma(a_i+1)} 
	\left(-\frac{\partial}{\partial \SP_i} \right) f(\SP)
	\label{eq:parametric-ibp}%
\end{equation}
shows that for any value of $i$ for which $a_i \leq 0$, the variable $\SP_i$ can be eliminated from \eqref{eq:parametric} by replacing the integrand $f(\SP)$ with $\left[(-\partial_i)^{-a_i} f(\alpha)\right]_{\SP_i = 0}$.
For our calculation, the remaining (undetermined by the constraints above) master integrals have at most 12 positive indices $a_i > 0$. Our approach was to compute these integrals by integrating out the (at most 12) remaining Schwinger parameters $\SP_i$ one by one.

This is possible with the algorithm \cite{Brown:TwoPoint} of Francis Brown, which uses a class of special functions called hyperlogarithms to express all intermediate results as iterated integrals.\footnote{%
	A slightly different, multivariate approach \cite{BognerBrown:SymbolicIntegration,BognerBrown:GenusZero} was implemented in \cite{Bogner:MPL}.
}
The output of this method are special values of multiple polylogarithms, and specifically in our case we obtain rational linear combinations of multiple zeta values (MZVs),
\begin{equation}
	\mzv{n_1,\ldots,n_d}
	\defas
	\sum_{1 \leq k_1<\cdots < k_d} \frac{1}{k_1^{n_1}\! \cdots k_d^{n_d}}
	,
	\label{eq:mzv}%
\end{equation}
which are (conjecturally) transcendental numbers; indexed by integers $n \in \NN^d$ with $n_d \geq 2$, and generalize the Riemann zeta values (the case $d=1)$. We call $d$ the depth and $n_1+\cdots+n_d$ the (transcendental) weight.

To perform the actual calculations, we used the implementation {\HyperInt} \cite{Panzer:2014caa}. In order to apply this tool, a technical condition called linear reducibility has to be fulfilled by the graph polynomials $\U$ and $\F$, which turns out to be the case for the integrals under consideration.
Then the only remaining obstacle, apart from practical limits to runtime and memory, are divergences.
The reason is that the parametric integration with hyperlogarithms can only be applied to integrands that comprise rational functions and polylogarithms, so we must expand the integrand \eqref{eq:parametric} in $\epsilon$, prior to integration.
But the two-point integrals under consideration typically have divergences which are only regulated by $\epsilon$, such that the $\epsilon$-expansion of the integrand usually results in divergent, ill-defined integrals.

One way to avoid this problem is by choosing the basis of master integrals in the first place in such a way that they are all finite---free of any (sub-)divergences. In principle, this is always possible \cite{Panzer:DivergencesManyScales,ManteuffelPanzerSchabinger:QuasiFinite} and feasible via IBP reductions \cite{ManteuffelPanzerSchabinger:FF}. In our problem, however, the enormous complexity of the IBP reductions makes a transformation to a finite basis seem a daunting task.
However, this is not necessary, since we already obtained the vast majority of master integrals using the relations mentioned earlier.

Given any finite integral, if we can manage to reduce it to our basis of master integrals, its $\epsilon$-expansion coefficients will be linearly related to the $\epsilon$-expansions of the master integrals. Computing fifteen finite integrals with {\HyperInt}, we were thus able to obtain more coefficients in the $\epsilon$-expansions of the master integrals.

In one case, we computed an integral with a single one-loop subdivergence.
After integrating out the subdivergence, we are left with an integral where one of the indices $a_i = n\epsilon$ is proportional to $\epsilon$. In this situation, we first integrate out all of the other Schwinger parameters to obtain the coefficients of the final integrand $f(\SP_i) = \sum_{k \geq 0} \epsilon^k f_k(\SP_i)$ as hyperlogarithms $f_k(\SP_i)$ in $\SP_i$, and then use integration by parts, as in\eqref{eq:parametric-ibp}, to rewrite the last integration as
\begin{equation*}
	\int_0^{\infty} \SP_i^{n\epsilon-1} f(\SP_i) \ \dd \SP_i
	= 
	-\frac{1}{n\epsilon}
	\sum_{j,k \geq 0} \epsilon^{j+k}
	\int_0^{\infty} \frac{(n \log \SP_i)^j}{j!} f_k'(\SP_i) \ \dd \SP_i
	.
\end{equation*}
Note that the subdivergence has been absorbed in the explicit pole $1/(n\epsilon)$ as a prefactor, and the parametric integral over $\SP_i$ on the right-hand side is convergent.

\section{Checks of results}
\label{sec:checks}

Some master integrals are in fact quite easy to evaluate, like the integrals which are products of the known lower-loop master integrals (we used these inputs as explained above).

On the other hand, there are several integrals for which some of the integrations can be performed explicitely, but we purposefully left them undetermined in the master bootstrap so that they could serve as a non-trivial check of our method. Whenever an integration variable appears only in two of the propagators, we can use the one-loop formula
\begin{equation}\label{oneloopformula}
	\int\frac{\dd[d]x_5}{(x_{58}^2)^a (x_{59}^2)^b}
	= \frac{G(a,b)}{\epsilon \,G(1,1)}\frac{1}{(x_{89}^2)^{a+b-d/2}}\,,
\end{equation}
in terms of the $G$-function which is defined as
\begin{equation}
	G(a,b)
	= \frac{
		\Gamma(a+b-d/2)
		\Gamma(d/2-a)
		\Gamma(d/2-b)
	}{
		\Gamma(a)
		\Gamma(b)
		\Gamma(d-a-b)
	}\,.
	\label{eq:G}%
\end{equation}
Note that we interpret all integrals, like \eqref{oneloopformula}, in the $G$-scheme from \cite{Chetyrkin:1980pr}. This means that each $d$-dimensional integration measure $\dd[d] x_i$ is understood with an addidional absorbed factor of $\left[ \pi^{d/2} \epsilon G(1,1)\right]^{-1} = \pi^{\epsilon-2} \Gamma(2-2\epsilon)/\left[ \Gamma^2(1-\epsilon) \Gamma(1+\epsilon)\right]$, in comparison with the standard Lebesgue measure. In particular, this normalizes the bubble integral to
\begin{equation*}
	\int\frac{\dd[d]x_5}{x_{58}^2 x_{59}^2}
	= \frac{1}{\epsilon}\frac{1}{(x_{89}^2)^{\epsilon}}.
\end{equation*}
Ten of the genuine five-loop master integrals can be determined exactly through successive applications of formula \eqref{oneloopformula}, and these exact expressions agree with our $\epsilon$-expansions obtained through the bootstrap approach.

There are also fifteen master integrals with three trivial integrations, which are then reduced to two-loop integrals with non-integer exponents. One such example is the integral
\begin{equation*}
	I_1
	=\int \frac{\dd[4] x_1 \ldots \dd[4] x_5}{x_{04}^2\,x_{05}^2\, x_{46}^2 \,x_{56}^2 \,x_{14}^2 \,x_{15}^2 \,x_{24}^2 \, x_{25}^2 \,x_{34}^2\, x_{35}^2} 
	= \frac{ G(2,1)}{\epsilon^3 G(1,1)} F(1,1,1,1, 3 \epsilon)\,,
\end{equation*}
where we introduced the two-loop integral
\begin{equation*}
	F(1,1,1,1, a) 
	\defas \int \frac{\dd[4] x_1 \dd[4] x_2}{x_{01}^{2}\,x_{02}^{2}\,x_{16}^{2}\,x_{26}^{2}\,(x_{12}^{2})^{a}}\,.
\end{equation*}
However, as it will become clear below, it is useful to consider a slightly modified version of the master integral
\begin{equation*}
	I_2
	= \int \frac{\dd[4] x_1 \ldots \dd[4] x_5}{x_{04}^2\,x_{05}^2\, x_{46}^2 \,x_{56}^2 \,x_{14}^2 \,x_{15}^2 \,x_{24}^2 \, x_{25}^2 \,x_{34}^2\, (x_{35}^2)^2}
	= \frac{ G(2,1)}{\epsilon^3 G(1,1)} F(1,1,1,1,1+ 3 \epsilon)\,.
\end{equation*}
The resulting two-loop integral has been obtained before and is given by the following expansion of homogeneous transcendental weight \cite{Kotikov:1995cw,Bierenbaum:2003ud}:
\begin{align*}
\frac{1}{1-2\epsilon}  F(1,1,1& ,1, 1+3 \epsilon)
	=  6\, \zeta_3 
	+ \frac{\pi^4}{10}\,\epsilon 
	+ 312 \,\zeta_5\,\epsilon^2  
	+ \left(
		\frac{17 \,\pi^6}{21}
		-288 \,\zeta_3^2
	\right)  \epsilon^3
\nonumber\\
	& + \left(
		-\frac{48 \,\pi^4 \,\zeta_3}{5} 
		+ 11283 \,\zeta_7
	\right) \epsilon^4
	+\left(
		\frac{6971 \,\pi^8}{2100} 
		+ 1944\, \zeta_{3,5} 
		- 17100 \,\zeta_3 \zeta_5
	\right)  \epsilon^5
\nonumber\\
	&+ \left(
		-\frac{306 \,\pi^6 \zeta_3}{7}
		+ 7812\, \zeta_3^3 
		- 447 \,\pi^4 \zeta_5 
		+ 370374\, \zeta_9
	\right)\epsilon^6
	+ \mathcal{O}(\epsilon^7)\,.
\end{align*}
We have in this way determined the integral $I_2$ and can relate it to the original master integral through an IBP reduction
\begin{equation*}
	I_2
	= -\frac{(d-3)^2 (3d-10)(5d-16)(9d-34)}{4(2d-7)(d-4)^3} I_3
	+  \frac{(d-3)(5d-18)}{4(d-4)} I_1 \,.
\end{equation*}
All integrations of the new integral $I_3$ are easily obtained using equation \eqref{oneloopformula},
\begin{equation*}
	I_3 
	= \int\frac{\dd[4] x_1 \ldots \dd[4] x_5}{x_{05}^2 \,x_{26}^2\, x_{36}^2 \,x_{46}^2 \,x_{12}^2\, x_{13}^2\, x_{14}^2 \,x_{15}^2} 
	= \frac{G(3\epsilon, \epsilon)}{\epsilon^5 G(1,1)}\frac{1}{(x_{06})^{2-5\epsilon}}\,,
\end{equation*}
and we thus obtain the expansion of the master integral $I_1$.  It is a highly non-trivial test of our master bootstrap program that the results obtained in all cases match exactly the known analytic expressions. 

Finally, we performed a few further checks of the method proposed in the previous section. Several of the four-point integrals considered could be mapped through magic identities to the ladder diagram, which is known exactly  \cite{Usyukina:1993ch}, and so the result of the asymptotic expansions could be checked against the coincidence limit of the exact four-point integral. 
We also checked that our expansions match the results for $5$-loop $p$-integrals obtained from the $6$-loop $\phi^4$ $\beta$-function computation \cite{Kompaniets:2016hct,Kompaniets:2017yct}, and the master integrals obtained in this work were also used in \cite{Georgoudis:2017meq} to reproduce the five-loop anomalous dimension of the Konishi operator in $\mathcal{N}=4$ SYM. Finally, we were able to check the integrability predictions for massless two-point integrals presented in \cite[Eqs.~(73--74)]{Caetano:2016ydc}.\footnote{We were even able to fix their undetermined coefficients to be $d_1= \frac{1}{3}$ and $p_1= -\frac{293}{30} -\frac{37 \pi^2}{24}+ \frac{13 \pi^4}{90}-\frac{325 \zeta_3}{18}$.}

\section{Conclusion}

In this paper we have developed a method for determining the expansions of five-loop massless propagator-type integrals. The constraints imposed from finiteness and conformal symmetry of higher-point integrals can be easily generalized to higher loops, and so the bottleneck in this method becomes the ability to perform IBP reductions in an efficient way. 
Some of the integrals that were not determined through the linear system of constraint equations were linearly reducible and so we were able to use {\HyperInt} to evaluate them directly despite the high loop order.
	
While the results obtained here are for the expansions of master integrals around four dimensions, it is possible to use lowering or raising relations to obtain the expansions around any other even dimension \cite{Tarasov:1996br}. In order to find them for odd $d$ one would have to find new finite conformal integrals and bootstrap the master integrals from the new set of constraints obtained.
	
If the normalization of the integrals is the $G$-scheme as in \eqref{oneloopformula}, then we can make a few general remarks that resemble similar observations at lower loop orders:
\begin{enumerate}
	\item We find that the coefficients of the master integrals are rational linear combinations of the MZVs \eqref{eq:mzv}. If the terms of transcendental weight 9 appear first at order $p$ in $\epsilon$, then at order $\epsilon^{p-i}$ the transcendental weight never exceeds $9-i$. Furthermore, the coefficients with $9-i<3$ are always rational numbers, which means that $\zeta_2$ is absent from the expansions. 
	In fact, it was proven \cite[section~9.3]{Brown:FeynmanAmplitudesGalois} that transcendentals of weight one (i.e.\ logarithms) can never appear in any two-point graph; the absence of $\zeta_2$ remains open but is expected to follow from the same ``small graphs'' principle.

	\item The divergent part is constrained even further, as the coefficients of $\epsilon^{-n}$ never have transcendental contributions of weight larger than $9-2n$ (where $n>0$). In particular, the two highest poles $\epsilon^{-5}$ and $\epsilon^{-4}$ always have rational coefficients.

	\item It was observed in \cite{Baikov:2010hf} that an $\epsilon$-dependent transformation of the zeta functions eliminates even zetas from the expansions of four-loop two-point integrals. A careful analysis of our results shows that these transformations can be generalized so that at five loops we have only dependence on the following five linear combinations:
	\begin{equation}\label{eq:eps-zetas}\begin{split}
		\hat{\zeta}_3 
		&\defas \zeta_3 + \frac{3 \epsilon}{2} \zeta_4 - \frac{5 \epsilon^3}{2} \zeta_6 + \frac{21 \epsilon^5}{2} \zeta_8\,, 
		\qquad
		\hat{\zeta}_5 
		\defas \zeta_5 + \frac{5 \epsilon}{2} \zeta_6 - \frac{35 \epsilon^3}{4} \zeta_8\,,
		\\
		\hat{\zeta}_7 
		&\defas \zeta_7 + \frac{7 \epsilon}{2} \zeta_8 \,,
		\qquad
		\hat{\zeta}_{3,5} 
		\defas \mzv{2,6}- \mzv{5,3} -3\epsilon\, \zeta_4 \,\zeta_5 + \frac{5 \epsilon}{2} \zeta_3\, \zeta_6
		\qquad\text{and}\qquad
		\zeta_9.
	\end{split}\end{equation}
	We have also verified that such a transformation applies to the expansions of lower-loop two-point integrals  \cite{Lee:2011jt}. This implies that any finite linear combination of five- or lower-loop massless  two-point integrals $P_i(\epsilon)$ of the form 
	\begin{equation*}
		\sum_i c_i(\epsilon) P_i(\epsilon)\,,
	\end{equation*}
	cannot contain any even zeta in the limit of small $\epsilon$ if the expansion of the functions $c_i(\epsilon)$ is expressed only through rational coefficients. Furthermore, the transformation also applies to all momentum-space $p$-integrals from the $6$-loop $\phi^4$ computation \cite{Kompaniets:2016hct,Kompaniets:2017yct}.

	\item
	Note that, according to \eqref{eq:eps-zetas}, any dependence on multiple zeta values must come in the combination $\varpi=\mzv{2,6}- \mzv{5,3} \approx -0.1868414$. This particular combination has been observed already long ago in the periods of $\phi^4$ theory: In the census \cite{Schnetz:2008mp}, the only $6$-loop $\phi^4$ periods\footnote{These are equal to some finite, $5$-loop massless propagators.} that are not polynomials in Riemann zeta values are\footnote{%
	To compare, note that
	$
		\varpi
		= \frac{3}{5} \mzv{3,5} + \mzv{3} \mzv{5} - \frac{29}{20} \zeta_8
		= \frac{16}{9} N_{3,5} - \frac{1}{4} \mzv{3} \mzv{5}
	$
	in terms of
	$
		N_{3,5}
		=Q_8
	$
	defined in \cite{Schnetz:2008mp}.
}
	\begin{equation}
		P_{6,3} = 108 \mzv{3} \mzv{5} + 144 \varpi
		\quad\text{and}\quad
		P_{6,4} = -288 \mzv{3} \mzv{5} - 2304 \varpi
		.
		\label{eq:P63,P64}%
	\end{equation}
	Our calculation shows that this remains the only irreducible multiple zeta value up to six loops, even after leaving the restriction to $\phi^4$ graphs and considering all two-point integrals.\footnote{Beware, however, that some two-point integrals and several non-planar $p$-integrals were left undetermined in our calculation.}
	Furthermore, the double zeta value $\varpi$ has a peculiar number theoretic property: Within the theory of motivic periods \cite{Brown:NotesMotivicPeriods} (an extension of classical Galois theory to certain transcendental numbers), one can compute a coaction of the motivic counterpart of multiple zeta values. Following \cite{Brown:DecompositionMotivicMZV}, one gets
		\begin{equation}
			\Delta' \left(\mzv[\motivic]{2,6} - \mzv[\motivic]{5,3}\right)
			= \mzv[\deRham]{3} \otimes \mzv[\motivic]{5} - 2 \mzv[\deRham]{5} \otimes \mzv[\motivic]{3}.
			\label{eq:coaction}%
		\end{equation}
		What this means is that the only Galois conjugates of $\varpi$ (besides $1$ and $\varpi$ itself) are the odd Riemann zeta values $\mzv{3}$ and $\mzv{5}$. In contrast, most MZV at weight 8 have further Galois conjugates $\mzv{2}$ and $\mzv{2}\mzv{3}$. The absence of the latter from $\varpi$ can be seen as a far reaching extension of point 1.\ above: Not only does $\mzv{2}$ not appear by itself in the expansion of massless two-point functions, it is also not a Galois conjugate of any coefficient in the $\epsilon$-expansion.

		This observation is a basic case of the \textit{coaction principle} \cite{PanzerSchnetz:Phi4Coaction} observed in $\phi^4$ periods, which says that the rational linear combinations of the leading coefficients in the $\epsilon$-expansion are closed under the action of the motivic Galois group (this principle seems to apply to other physical observables as well, see \cite{Schnetz:CoactionElectron}).
\end{enumerate}

Propagator integrals are of crucial importance for the evaluation of higher-loop corrections of different physical observables. The results obtained here, for all planar and many of the non-planar five-loop master two-point integrals, open the possibility to tackle five-loop problems that were until now inaccessible, such as correlation functions and form factors in $\mathcal N=4$ SYM \cite{Georgoudis:2017meq,Yang:2016ear}.

\section*{Acknowledgments}
We thank Joe Minahan, Roman Lee, Alexander Smirnov, Yang Zhang and Oliver Schnetz for useful discussion and help. We are also indebted to Gregor K\"alin for help with the optimized installation of FIRE.
The computations were performed on resources provided by the Swedish National Infrastructure for Computing (SNIC)  at Uppmax and HPC2N.
We would like to thank Lars Viklund at HPC2N and Linus Nilsson at Uppmax for their assistance during the computations and for allowing us to run jobs over the time limit. 

V.G. is funded by FAPESP grant 2015/14796-7 and CERN/FIS-NUC/0045/2015. The work of A.G. is supported by the Knut and Alice Wallenberg Foundation under grant \# 2015-0083. A.G.  would like to thank FAPESP grant 2016/01343-7 for funding part of his visit to ICTP-SAIFR in March 2017 where part of this work was done. R.P.  was funded under VR2016-03503 and he was also supported by SFI grant 15/CDA/3472.

\appendix
\section{Master integrals in momentum space}

In this paper we obtained $\epsilon$-expansions of five-loop two-point integrals in position space. For some applications, however, it is more commonplace to work in momentum space, and thus we would like to determine $p$-integrals from our results.
If a $p$-integral is planar, the associated dual graph corresponds to a planar integral in position space, and our results can be translated in a straightforward way.

For non-planar integrals, however, the Fourier transform is the only connection between position- and momentum space. It transforms a power of a propagator $x^2$ according to
\begin{equation}\label{eq:Fourier}
	\Fourier{\frac{1}{x^{2 \lambda}}}
	= \int \frac{\dd[d]x}{\pi^{d/2}} \frac{e^{i p \cdot x}}{x^{2\lambda}}
	=
	\frac{ 4^{d/2-\lambda} \Gamma(d/2-\lambda)}{\Gamma(\lambda) }
	\frac{1}{(p^{2})^{d/2-\lambda}}
	,
\end{equation}
turning the integer exponents $\lambda=a_{ij}$ from \eqref{eq:pint} into $\epsilon$-dependent exponents $d/2-\lambda=2-\lambda-\epsilon$.
For example, let us consider our result\footnote{Note that it is proportional to the primitive $P_{6,2}$ of Schnetz's census \cite{Schnetz:2008mp}.}
for the non-planar master integral from Figure~\ref{fig:NPint},
\begin{figure}
	\centering
	\begin{tikzpicture}[scale=0.33,node/.style={draw,shape=circle,fill=black,scale=0.5}]
	\path (-9,0) node[label=\small $0$] (p0) {}
	(9,0) node[label=\small{$1$}] (p1) {}
	(0,12) node[label=\small $2$] (p2) {}
	(-8,8) node[label=\small $3$] (p3) {}
	(1,5) node[label=\small $4$] (p4) {}
	(0,0) node[label=\small $5$] (p5) {}
	(8,8) node[label=\small $6$] (p6) {};
	\draw[ thick]  (-8,8) -- (0,12) -- (8,8) --(8,0) -- (-8,0);
	\draw[ thick]  (8,8) -- (0,0) -- (-8,8) --(-8,0) -- (0,5)--(0,12);
	\draw[ thick]  (8,0) -- (0,5);
	\end{tikzpicture}
	\caption{A non-planar five-loop two-point integral. Its Fourier transforms is related to a non-planar master integral in momentum space.}%
	\label{fig:NPint}%
\end{figure}
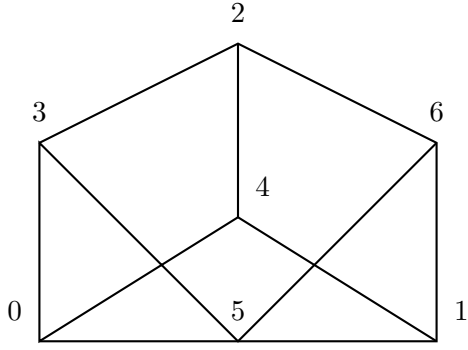
\begin{equation}
	M_x \defas
	\int \frac{\dd[d] x_1 \ldots \dd[d] x_5}{x_{03}^2\, x_{04}^2\,x_{05}^2\, x_{16}^2\, x_{26}^2\, x_{56}^2 \, x_{14}^2\,x_{15}^2\,x_{23}^2\,x_{24}^2\,x_{35}^2}
	= \frac{8 \zeta_3^3 + \frac{1063}{9} \zeta_9 + \mathcal{O}(\epsilon)}{(x_{06}^2)^{1+5 \epsilon}}\,.
	\label{eq:Mx}%
\end{equation}
Applying the Fourier transform \eqref{eq:Fourier} to the right-hand side, we find (to leading order in $\epsilon$)
\begin{equation}\label{FT1}
	\Fourier{M_x}
	= \frac{4}{p^2} \left(8 \zeta_3^3 + \frac{1063}{9} \zeta_9 \right)
	+ \mathcal{O}(\epsilon)
	\,.
\end{equation}
On the other hand, we can perform the Fourier transform at the level of the integrand of $M_x$, and thus obtain a momentum space representation of the integral. According to \eqref{eq:Fourier}, we have
\begin{equation}\label{FT2}
	\Fourier{M_x} = 4^{1-6\epsilon} \Gamma^{11}(1-\epsilon) M_p \,,
\end{equation}
with the $p$-integral $M_p$ (obtained by attaching external legs to $0$ and $6$ in Figure~\ref{fig:NPint}) given by
\begin{equation*}
 	M_p \defas 
	\int \frac{\dd[d] k_1 \cdots \dd[d] k_5}{\left[ k_1^{2} k_3^{2} k_4^{2} (p-k_2)^{2} (p-k_5)^{2} (p+k_{12}+k_{45}-k_3)^{2} k_{12}^{2} k_{13}^{2} k_{24}^{2} k_{45}^{2} (k_{12}-k_{34})^{2}\right]^{1-\epsilon}}
	.
\end{equation*}
Since this integral is finite, we can safely extract the leading term of its $\epsilon$-expansion by setting all propagator exponents in the integrand to one, in which case $M_p$ becomes a non-planar master $p$-integral in momentum space. Therefore, equating \eqref{FT1} with \eqref{FT2} produces the leading term of the $\epsilon$-expansion of $M_p$,
\begin{equation}
	M_p 
	= \frac{1}{p^2} \left(8 \zeta_3^3 + \frac{1063}{9} \zeta_9 \right)
	+ \mathcal{O}(\epsilon)
	\,.
\end{equation}
We have performed this analysis for all position-space integrals with eleven denominators, and in that way we were able to extract the leading (finite) order of 20 non-planar master integrals in momentum space. 

Note that the reverse approach was used in \cite{Eden:2012fe}, where the authors used the Fourier transform to obtain non-planar position-space integrals from known momentum-space integrals.

\bibliographystyle{unsrt}
\bibliography{biblio}

\end{document}